\pdfoutput=1
\documentclass[
	physrev,
	reprint,
	amsmath,
	amsfonts,
	]{revtex4-2}

\usepackage[all]{nowidow}
\usepackage{csquotes}

\newcommand*{\latin}[1]{#1}

\newcommand*{\ie}{\latin{i.\,e.}}

\newcommand*{\computerprogram}[1]{\textsc{#1}}

\DeclareMathOperator{\asinh}{asinh}

\DeclareMathOperator{\atanh}{atanh}

\newcommand*{\increment}{\mathrm{\Delta}}
\newcommand*{\differential}[1]{\mathinner{\mathrm{d}#1}}
\newcommand*{\archimedesconstant}{\mathrm{\pi}}
\newcommand*{\eulersnumber}{\mathrm{e}}

\newcommand*{\Au}{\textsuperscript{197}Au}
\newcommand*{\Pb}{\textsuperscript{208}Pb}

\newcommand*{\N}{N_{p-\bar{p}}}
\newcommand*{\dNdy}{\frac{\differential{\N}}{\differential{y}}}
\newcommand*{\dNdyinline}{{\differential{\N}}/{\differential{y}}}
\newcommand*{\pdfxy}{f}
\newcommand*{\pdfy}{\psi}
\newcommand*{\pdfp}{\phi}

\newcommand*{\lorentzfactor}{\gamma}
\newcommand*{\azimuthalangle}{\varphi}

\newcommand*{\referenceframe}{\mathfrak{F}}

\newcommand*{\fermi}{_{\mathrm{F}}}
\newcommand*{\beam}{_{\mathrm{b}}}
\newcommand*{\sNN}{s_{\mathrm{NN}}}

\newcommand*{\initial}{_{\mathrm{i}}}
\newcommand*{\final}{_{\mathrm{f}}}
\newcommand*{\stationary}{_{\mathrm{s}}}
\newcommand*{\cgc}{_{\mathrm{CGC}}}
\newcommand*{\ti}{t\initial}
\newcommand*{\tf}{t\final}

\newcommand*{\drift}{\mu}
\newcommand*{\diffusion}{\sigma}
\newcommand*{\diffusivity}{D}
\newcommand*{\timespan}{\increment t}
\newcommand*{\evolutionscale}{\diffusivity \timespan}

\newcommand*{\gssexponent}{\lambda}
\newcommand*{\gssprefactor}{Q_0^2}

\newcommand*{\ndf}{\mathrm{ndf}}

\usepackage{mathtools}
\DeclarePairedDelimiter{\parens}{\lparen}{\rparen}
\DeclarePairedDelimiter{\bracks}{\lbrack}{\rbrack}

\DeclarePairedDelimiter{\verts}{\lvert}{\rvert}

\usepackage[detect-all]{siunitx}
\usepackage{dcolumn}

\begin{document}

\title{Baryon stopping as a relativistic Markov process in phase space}

\author{Johannes Hoelck}
\author{Georg Wolschin}
\email{wolschin@thphys.uni-heidelberg.de}
\affiliation{Institute for Theoretical Physics, Heidelberg University, Philosophenweg 12--16, 69120 Heidelberg, Germany, EU}

\date{\today}

\begin{abstract}
We reconsider baryon stopping in relativistic heavy-ion collisions in a nonequilibrium-statistical framework.
The approach combines earlier formulations based on quantum chromodynamics with a relativistic diffusion model through a suitably derived fluctuation--dissipation relation, thus allowing for a fully time-dependent theory that is consistent with QCD.
We use an existing framework for relativistic stochastic processes in spacetime that are Markovian in phase space, and adapt it to derive a Fokker--Planck equation in rapidity space, which is solved numerically.
The time evolution of the net-proton distribution function in rapidity space agrees with stopping data from the CERN Super Proton Synchrotron and the BNL Relativistic Heavy Ion Collider.
\end{abstract}

\maketitle

\section{\label{sec:introduction}Introduction}
In relativistic heavy-ion collisions at the CERN Super Proton Synchrotron (SPS), the BNL Relativistic Heavy Ion Collider (RHIC), or the CERN Large Hadron Collider (LHC), the incoming baryons are being slowed down (\enquote{stopped}) as they interpenetrate each other, while in the spatial region between the receding, highly Lorentz-contracted fragments \cite{Bjorken-1983-Phys.Rev.D27} a hot fireball is formed, which cools during its expansion and eventually hadronizes in a parton--hadron crossover.
Of particular interest is the initial stage of such a collision with the local thermalization of quarks and gluons, and the simultaneous stopping of the baryons.
The latter occurs essentially through collisions of the incoming valence quarks with soft gluons in the respective other nucleus.

Various models to account for the stopping process and its energy dependence have been developed, for example, in Refs.\,\cite{MehtarTaniWolschin-2009-Phys.Rev.Lett.102,MehtarTaniWolschin-2009-Phys.Rev.C80,DuraesGianniniGoncalvesNavarra-2014-Phys.Rev.C89} and related works, which are relying on the appropriate parton distribution functions and hence on quantum chromodynamics (QCD).
These models yield agreement with the available stopping data at SPS and RHIC, such as the distributions of net protons (protons minus produced antiprotons) in longitudinal
rapidity space, and also provide predictions at LHC energies, where stopping data at forward rapidities are not yet available.
They do not, however, provide the time development from the initial distribution at the instant of the collision to the final, measured one.

Complementary time-dependent approaches to stopping and local equilibration have relied on phenomenological, nonequilibrium-statistical approaches:
A linear Fokker--Planck equation for the net-baryon rapidity distribution function had been proposed in Ref.\,\cite{Wolschin-1999-Eur.Phys.J.A5}, which accounts for the time evolution of the net-baryon or net-proton rapidity distributions in a two-source relativistic diffusion model (RDM).
Variants of the model with a nonlinearity in the diffusion term have subsequently been suggested in Refs.\,\cite{Lavagno-2002-PhysicaA305,RybczynskiWlodarczykWilk-2003-Nucl.Phys.BProc.Suppl.122,AlbericoLavagno-2009-Eur.Phys.J.A40} and related works, which assume, however, the debatable validity \cite{Wolschin-2016-Phys.Rev.C94} of nonextensive statistics.
A linear diffusion model for particle production was also put forward in Ref.\,\cite{BiyajimaIdeMizoguchiSuzuki-2002-Prog.Theor.Phys.108} and subsequent works.
For produced particles at RHIC and LHC energies, a third (midrapidity) source is essential to cover pair production processes in the central fireball that provide the bulk of charged-hadron generation at sufficiently high energy \cite{Wolschin-2016-Phys.Rev.C94}.
When considering net baryons or protons, however, the midrapidity source cancels out because it is equally composed of particles and antiparticles.
%
The highly nonlinear local thermalization of quarks and gluons in the initial stages of the collision can be modeled through quantum Boltzmann-like collision terms, which require numerical solutions \cite{BlaizotLiaoMcLerran-2013-Nucl.Phys.A920,BlaizotLiaoMehtarTani-2017-Nucl.Phys.A961}, but also a schematic model has been developed that accounts for the fast local equilibration through analytical solutions of a nonlinear partial differential equation \cite{Wolschin-2020-Universe6}.

Diffusion models are being used in many areas of physics, chemistry, and biology \cite{Jost-1960,Crank-1979}.
They have originally been developed by Einstein and Smoluchowski to provide a mesoscopic theory of Brownian motion~\cite{Einstein-1905-Ann.Phys.322,Smoluchowski-1906-Ann.Phys.326,Smoluchowski-1916-Ann.Phys.353} as linear differential equations for the Brownian particles' single-particle distribution function.
Alternative microscopic approaches treat the Brownian particles' trajectories as stochastic processes in position space, for example in the form of a Wiener~\cite{Wiener-1923-J.Math.Phys.2} or Ornstein--Uhlenbeck process~\cite{UhlenbeckOrnstein-1930-Phys.Rev.36}.
Stochastic processes and stochastic differential equations have subsequently been considered in more generality in a newly established branch of mathematics, stochastic calculus, with notable contributions by It{\^o}~\cite{Ito-1944-Proc.Imp.Acad.20,Ito-1946-Proc.Jpn.Acad.22}, Stratonovich~\cite{Stratonovich-1964-VestnikMoskov.Univ.Ser.IMat.Mekh.1}, Fisk~\cite{Fisk-1965-Trans.Amer.Math.Soc.120}, and Klimontovich~\cite{Klimontovich-1990-PhysicaA163} who introduced various concepts of a stochastic integral, each with different mathematical properties and physical interpretations.
In some cases, connections between the micro- and mesoscopic formulation can be established through a Kramers--Moyal expansion~\cite{Kramers-1940-Physica7,Moyal-1949-J.R.Stat.Soc.B11} or the Feynman--Kac formula~\cite{Kac-1949-Trans.Amer.Math.Soc.65}.

Following the discovery of special relativity, it became clear that statistical physics in general, and diffusion models in particular, would have to be adapted \cite{DebbaschRivetLeeuwen-2001-PhysicaA301} to meet the requirements imposed by a limited velocity of light.
Especially, nontrivial Lorentz-invariant stochastic processes for spacetime coordinates are necessarily non-Markovian \cite{Lopuszanski-1953-ActaPhys.Polon.12,Dudley-1966-Ark.Mat.6,Hakim-1968-J.Math.Phys.9}, which makes a straightforward generalization of the nonrelativistic diffusion equation to special relativity impossible.
While a general stochastic process may depend on any finite or infinite number of its prior realizations, Markov processes~\cite{Markov-1906-Izv.Fiz.Matem.Obsch.Kazan.Univ.15} have no memory of the past apart from their current state, which greatly simplifies their mathematical treatment.
Consequently, stochastic processes used in physical models often have the Markov property, in spite of being mathematically the exception rather than the rule.

As long as a process' memory is finite, \ie, only a finite number of its previous realizations affect the next value, it can be reformulated as a coupled system of multiple Markov processes through the introduction of additional variables~\cite{Kampen-1998-Braz.J.Phys.28}.
This has been used in Refs.\,\cite{DebbaschMallickRivet-1997-J.Stat.Phys.88,DunkelHaenggi-2005-Phys.Rev.E71,DunkelHaenggi-2005-Phys.Rev.E72,Zygadlo-2005-Phys.Lett.A345,DunkelHaenggiWeber-2009-Phys.Rev.E79,DunkelHaenggi-2009-Phys.Rep.471} to formulate relativistic phase-space diffusion processes based on a generalized Ornstein--Uhlenbeck process for the Brownian particle's momentum.
These processes are Markovian in phase space but lose the Markov property when expressed solely in spacetime coordinates.
The authors also deduced associated relativistic Kramers and Fokker--Planck equations for the particles' phase-space and momentum distribution functions, and derived fluctuation--dissipation relations suitable for an isotropic thermal background.

In the present work, we aim to derive a nonequilibrium-statistical diffusion model for baryon stopping in rapidity space that is based on the key premises of the phenomenological RDM, but is constructed from a consistent approach with relativistic Markov processes in phase space and incorporates the QCD-based theory through a suitably adapted fluctuation--dissipation relation.
The corresponding Fokker--Planck equation will enable us to account for the time evolution of the initial distribution functions from the onset of a relativistic heavy-ion collision to the final, measured distributions of net baryons or net protons in agreement with the available SPS and RHIC data.

The key assumptions for our nonequilibrium-statistical approach to stopping in relativistic heavy-ion collisions are presented in the next section, followed by the equations of motion in Langevin and Fokker--Planck formulation in Sec.\,\ref{sec:time-evolution}.
Drift and diffusion terms are discussed in Sec.\,\ref{sec:drift+diffusion}, as well as the expected stationary state derived from the earlier QCD formulation that allows us to formulate an appropriate fluctuation--dissipation relation that determines the course of the time evolution from the initial to the final net-proton rapidity distribution functions.
The latter are compared with available stopping data from SPS and RHIC experiments in Sec.\,\ref{sec:results}.
The conclusions are drawn in Sec.\,\ref{sec:conclusions}.

\section{\label{sec:net-protons}Net-proton rapidity spectra}
We model baryon stopping as a diffusive process in rapidity space of the participating nucleons whose dynamics are governed by a---not necessarily thermalized---fluctuating background representing the quarks and gluons of the fragments.
In this process, interactions between the partons and the background are assumed to prevail such that
the nucleon distribution function reduces to a superposition of single-particle probability density functions.
As the distribution functions of participant baryons are experimentally inaccessible, we consider only participant protons in our model and compare our result to the measured net-proton number density in rapidity space, \ie, the difference between the distributions of protons and antiprotons produced in the collision, which we expect to be reasonably close to the participant-proton distribution function.

To incorporate the spatial separation of the two nuclear fragments, we use a two-source ansatz~\cite{Wolschin-1999-Eur.Phys.J.A5} and completely disconnect the time evolution of particles originating from the forward- and backward-moving fragment through separate probability densities and fluctuation--dissipation relations.
Taking advantage of the symmetry of the system with respect to its center of momentum, we then write the net-proton number density in rapidity space~$\dNdyinline$ in the system's center-of-momentum frame~$\referenceframe$ as the superposition
\begin{equation}
	\dNdy(t;y) \approx \frac{\N}{2} \bracks*{\pdfy(t;+y) + \pdfy(t;-y)}
	\label{eq:dNdy}%
	\mathinner{,}
\end{equation}
where $\N$ denotes the net-proton number and $\pdfy(t;\pm y) \differential{y}$ the probability to find a participant proton from the forward- or backward-moving fragment, respectively, at time~$t$ with rapidity in $\bracks{y,y+\differential{y}}$.
%

\subsection{Initial state}
Prior to the collision of the nuclei at some time~$\ti$, we assume the system to be in an initial state where each nucleus can be approximated by a zero-temperature Fermi gas with appropriate Fermi momentum~$p\fermi$.
Then, the protons of each nucleus are distributed in the nucleus's rest frame~$\referenceframe_*$ according to the momentum-space probability density function
\begin{equation}
	\pdfp\initial(\vec{p}_*) = \frac{3}{4\archimedesconstant p\fermi^3} \mathinner{\Theta(p\fermi - \verts{\vec{p}_*})}
	\mathinner{,}
\end{equation}
which is given by a Heaviside step function~$\Theta$ scaled by a normalizing factor.
We determine the Fermi momentum through a simple potential well model,
\footnote{We use a natural system of units where the speed of light in vacuum~$c$ and the reduced Planck constant~$\hbar$ are equal to unity, $c = \hbar = 1$.}
\begin{equation}
	p\fermi = \sqrt[3]{3\archimedesconstant^2 \frac{Z}{V_*}}
	\mathinner{,}\qquad
	V_* = \frac{4\archimedesconstant}{3} r_*^3
	\mathinner{.}
\end{equation}
Here, $Z$ denotes the nucleus's proton number, $V_*$ the nuclear charge volume, and $r_*$ the nuclear charge radius, which we take from Ref.\,\cite{AngeliMarinova-2013-At.DataNucl.DataTables99}.

\begin{figure}
	\includegraphics{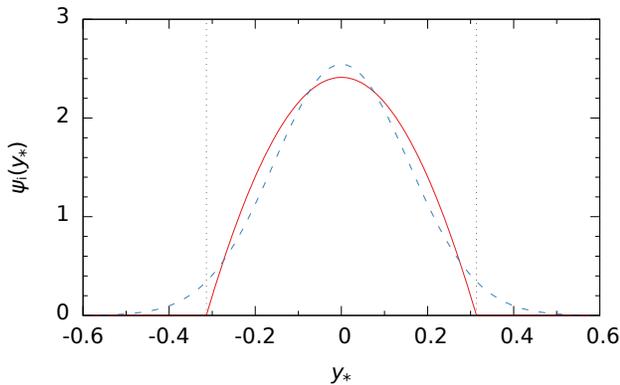}%
	\caption{\label{fig:pdf-initial}%
		Marginal rapidity probability density function~$\pdfy\initial$ (solid, red) of the participant protons in a \Pb~nucleus prior to the initial collision.
		For comparison, a normal distribution with zero mean and standard deviation $y\fermi/2$ is depicted as dashed blue curve.
		Dotted vertical lines indicate the Fermi rapidity~$\pm y\fermi$.
	}
\end{figure}

Choosing the orientation of $\referenceframe_*$ such that $p_*^3$ is parallel to the beam axis, we define the cylindrical coordinates
\begin{subequations}
\begin{gather}
	\lorentzfactor_{\perp*} = \sqrt{1 + (p_*^1/m)^2 + (p_*^2/m)^2}
	\mathinner{,}\\
	\azimuthalangle_* = \arctan(p_*^2/p_*^1)
	\mathinner{,}\\
	y_* = \atanh(p_*^3/p_*^0)
	\mathinner{,}
\end{gather}
\end{subequations}
with the proton mass~$m$.
Boosting to $\referenceframe$ leaves the transverse degrees of freedom unaffected ($\lorentzfactor_{\perp} = \lorentzfactor_{\perp*}$, $\azimuthalangle = \azimuthalangle_*$) while the longitudinal rapidity coordinate is shifted by the beam rapidity~$y\beam$, $y = y_* \pm y\beam$.
Integrating out $\lorentzfactor_{\perp}$ and $\azimuthalangle$, the initial rapidity-space probability density in $\referenceframe$, $\pdfy\initial(y) \equiv \pdfy(\ti;y)$, is found to be
\begin{multline}
	\pdfy\initial(y_* \pm y\beam) = \frac{1}{2} \sinh(y\fermi)^{-3} \mathinner{\Theta(y\fermi - \verts{y_*})}
	\\
	\times \cosh(y_*) \bracks*{\parens*{\frac{\cosh(y\fermi)}{\cosh(y_*)}}^3 - 1}
\end{multline}
with the Fermi rapidity~$y\fermi = \asinh(p\fermi/m)$.
Fig.\,\ref{fig:pdf-initial} shows $\pdfy\initial$ as a function of $y_*$ for the isotope \Pb.
The numerical values of $y\fermi$ for {\Au} and {\Pb} are \num{0.3134} and \num{0.3136}, respectively, and differ only slightly in the fourth decimal place.

A more realistic description of the initial state is principally desirable (for example, with a finite temperature); however, the exact form of the initial probability density function hardly influences the later stages of the time evolution.
Hence, normal or delta distributions are often used as convenient approximations for the initial state \cite{ForndranWolschin-2017-Eur.Phys.J.A53}.

\subsection{Final state}
For $t > \ti$, the system evolves in time, driven by the fluctuating background, until it reaches a final state at some time~$\tf$ when the partonic interactions between the nuclei effectively cease due to their increasing spatial distance.
Consequently, for comparison with experimental data from SPS and RHIC, we evaluate Eq.\,(\ref{eq:dNdy}) at $t = \tf$.
As we will see in Sec.\,\ref{sec:results}, the concrete value of $\tf$ is not important at this stage of the model, since it does only appear in products with other \latin{a priori} unknown quantities; it is not an observable.

\section{\label{sec:time-evolution}Time evolution}
The time evolution of the system will ultimately be governed by a Fokker--Planck equation for the single-particle probability density function~$\pdfy$, whose drift and diffusion coefficient functions will be derived in Sec.\,\ref{sec:drift+diffusion} from the expected mesoscopic behavior.
A similar evolution equation had previously been determined on a phenomenological level in the RDM~\cite{Wolschin-2016-Phys.Rev.C94} from comparisons with available SPS and RHIC data.
In this work, we will derive it from the underlying particle dynamics to shed light on the different assumptions entering our model.

\subsection{Langevin formulation}
We describe the trajectories of the individual protons as relativistic stochastic processes in spacetime that are Markovian when expressed in phase-space coordinates~\cite{Kampen-1998-Braz.J.Phys.28,DunkelTalknerHaenggi-2007-Phys.Rev.D75}.
In the following, these stochastic processes will be designated with uppercase letters, while lowercase letters denote the corresponding coordinates in the system's center-of-momentum frame~$\referenceframe$.
The equations of motion for the spacetime position~$X^\alpha(t)$ and energy and momentum~$P^\alpha(t)$ as a function of time~$t$ follow from a relativistic generalization~\cite{DebbaschMallickRivet-1997-J.Stat.Phys.88,DunkelHaenggi-2005-Phys.Rev.E71,DunkelHaenggi-2005-Phys.Rev.E72,Zygadlo-2005-Phys.Lett.A345,DunkelHaenggiWeber-2009-Phys.Rev.E79,DunkelHaenggi-2009-Phys.Rep.471} of the Ornstein--Uhlenbeck process~\cite{UhlenbeckOrnstein-1930-Phys.Rev.36} in phase space,
\begin{subequations}
\begin{gather}
	\differential{X^\alpha} = P^\alpha/P^0 \differential{t}
	\mathinner{,}\\
	\differential{P^i} = \drift_{p^i} \differential{t} + \sum_{k} \diffusion_{p^i\mkern-2mu,k} \differential{W_k}
	\mathinner{,}
\end{gather}
\label{eq:sde-position+momentum}%
\end{subequations}
where the Greek index $\alpha$ runs from 0 to 3 and the Latin indices $i,k$ from 1 to 3.
The momenta $P^i$ are driven by three independent standard Wiener processes~$W_k$ \cite{Wiener-1923-J.Math.Phys.2} representing the fluctuating background, while the particle energy~$P^0$ is fixed by the mass-shell condition, $P^0 = \sqrt{m^2 + \vec{P}^2}$, where $m$ denotes the proton mass.

The interaction between particles and background is governed by the drift coefficients~$\drift_{p^i}$ and diffusion coefficients~$\diffusion_{p^i\mkern-2mu,k}$:
The former represent directed, deterministic effects (for example, friction or pressure gradients) and determine the mean value of the stochastic process; the latter are connected to undirected, stochastic interactions (such as random particle collisions) and its variance.
In general, they can be functions of all involved stochastic processes $X^\alpha$ and $P^i$.
Here, however, we will assume that they depend on the momentum processes only.

If we let the 3-direction of the coordinate system coincide with the beam axis, we can replace $P^3$ with the stochastic process
\begin{gather}
	Y = \atanh(P^3/P^0)
	\mathinner{,}
\end{gather}
which corresponds to
the (longitudinal) rapidity~$y$.
The associated drift coefficient~$\drift_y$ and diffusion coefficients~$\diffusion_{y,k}$ can be related to $\drift_{p^i}$ and $\diffusion_{p^i\mkern-2mu,k}$ through differential calculus.

To simplify the following computations, we will assume that the longitudinal drift and diffusion coefficients' dependence on the transverse degrees of freedom is negligible, \ie, $\partial_{p^i} \drift_y = \partial_{p^i} \diffusion_{y,3} = 0$ and $\diffusion_{p^i\mkern-2mu,3} = \diffusion_{y,k} = 0$ for $i,k = 1,2$.
Then, the Langevin equations for $Y$ decouple from those of $P^1$ and $P^2$,
\begin{subequations}
\begin{gather}
	\differential{X^3} = \tanh(Y) \differential{t}
	\mathinner{,}\\
	\differential{Y} = \drift_y(Y) \differential{t} + \diffusion_{y,3}(Y) \differential{W_3}
	\label{eq:sde-rapidity}%
	\mathinner{.}
\end{gather}
\label{eq:sde-longitudinal}%
\end{subequations}

Further, we want to treat $\diffusion_{y,3}$ as a constant with respect to rapidity for now, since the nonconstant case entails some technical subtleties regarding discretization and interpretation of the Langevin equations~\cite{Ito-1944-Proc.Imp.Acad.20,Ito-1946-Proc.Jpn.Acad.22,Stratonovich-1964-VestnikMoskov.Univ.Ser.IMat.Mekh.1,Fisk-1965-Trans.Amer.Math.Soc.120,Klimontovich-1990-PhysicaA163,Kampen-1981-J.Stat.Phys.24}.
We intend, however, to address this issue in a forthcoming publication.

The choice of a constant diffusion coefficient is permissible here because the rapidity~$Y$ may assume any real value, and hence arbitrarily large changes by the driving Wiener process still result in a physically permissible state of $Y$.
By contrast, if we were to formulate a stochastic process for the particle's velocity or Lorentz factor, the diffusion coefficient would necessarily have to be nonconstant to prevent superluminal motion by suppressing fluctuations that would lead the stochastic process to an unphysical state~\cite{Kampen-1986-J.Stat.Phys.44}.

\subsection{Fokker--Planck formulation}
To obtain an equation for the time evolution of the single-particle probability density function associated with the particle trajectories discussed in the preceding section, we perform a Kramers--Moyal expansion \cite{Kramers-1940-Physica7,Moyal-1949-J.R.Stat.Soc.B11} with respect to the longitudinal stochastic processes defined in Eqs.\,(\ref{eq:sde-longitudinal}) and the transverse stochastic processes from Eqs.\,(\ref{eq:sde-position+momentum}).
As we have decoupled $X^3$ and $Y$ from the other processes, we can immediately integrate out the transverse coordinates $x^1$, $x^2$ and $p^1$, $p^2$, which leaves us with the Kramers equation for the marginal probability density function~$f$ of longitudinal position~$x^3$ and rapidity~$y$,
\begin{equation}
	\bracks*{\partial_t + \tanh(y) \partial_{x^3} + \partial_y \drift(y) - \tfrac{1}{2} \diffusion^2 \partial_y^2} \pdfxy(t;x^3,y) = 0
	\mathinner{,}
	\label{eq:ke-longitudinal}%
\end{equation}
with $\pdfxy(t;x^3\mkern-2mu,y) \differential{x^3} \differential{y}$ giving the probability to find a participant proton at time~$t$ with $X^3 \in \bracks{x^3,x^3+\differential{x^3}}$ and $Y \in \bracks{y,y+\differential{y}}$ in $\referenceframe$.
To ease notation, we drop the subscripts of the longitudinal drift and diffusion coefficients from now on as they are the only coefficient functions left.

Given an appropriate initial condition, we could in principle solve Eq.\,(\ref{eq:ke-longitudinal}).
However, since the position coordinate~$x^3$ is unobservable, we integrate it out, thus reducing Eq.\,(\ref{eq:ke-longitudinal}) to a Fokker--Planck equation for the marginal rapidity probability density~$\pdfy$,
\begin{gather}
	\partial_t \pdfy(t;y) = -\partial_y \bracks*{\drift(y) \pdfy(t;y)} + \tfrac{1}{2} \diffusion^2 \partial_y^2 \pdfy(t;y)
	\label{eq:fpe}%
	\mathinner{,}\\
	\pdfy(t;y) = \int \differential{x^3} \pdfxy(t;x^3,y)
	\label{eq:pdf-marginal}%
	\mathinner{,}
\end{gather}
where we have used that $\pdfxy$ must vanish at the boundaries and that $\drift$ and $\diffusion$ were assumed to be independent of $x^3$.
Alternatively, Eq.\,(\ref{eq:fpe}) can be rewritten as a continuity equation
\begin{equation}
	\partial_t \pdfy(t;y) + \partial_y j(t;y) = 0
\end{equation}
with the probability current density
\begin{equation}
	j(t;y) = \bracks*{\drift(y) - \tfrac{1}{2} \diffusion^2 \partial_y} \pdfy(t;y)
	\mathinner{,}
\end{equation}
which can be decomposed into an advective~($j_{\mathrm{a}}$) and a diffusive part~($j_{\mathrm{d}}$),
\begin{subequations}
\begin{align}
	j_{\mathrm{a}}(t;y) &= \drift(y) \pdfy(t;y)
	\mathinner{,}\\
	j_{\mathrm{d}}(t;y) &= - \tfrac{1}{2} \diffusion^2 \partial_y \pdfy(t;y)
	\mathinner{.}
\end{align}
\end{subequations}
In this context, the prefactor $\diffusion^2 / 2$ can be understood as the protons' diffusivity~$\diffusivity$ in rapidity space.

When defining nonlocal observables as in Eq.\,(\ref{eq:pdf-marginal}) in relativistic statistical physics, care has to be taken \cite{DunkelHaenggiHilbert-2009-Nat.Phys.5} because the involved integral introduces a dependence on the chosen hypersurface.
In our case, integration is done with respect to isochronous hypersurfaces in $\referenceframe$; we expect this to give a reasonable representation of the $\dNdyinline$ measuring process.
A more accurate treatment would require precise knowledge of the particle positions and detector layout, which is beyond the scope of this model.

\section{\label{sec:drift+diffusion}Drift and diffusion}
So far, we have left open the exact form of the drift and diffusion coefficients, apart from setting the latter constant with respect to rapidity.
Instead of deriving them from microscopic considerations, we will set the coefficients in a way that the solutions of the Fokker--Planck equation~(\ref{eq:fpe}) reproduce a certain expected mesoscopic behavior of the physical system to be modeled, as proposed in Refs.\,\cite{DebbaschMallickRivet-1997-J.Stat.Phys.88,DunkelHaenggi-2009-Phys.Rep.471}.
Possible choices include presetting the system's stationary state or specifying the time evolution of some macroscopic observable.
Generally, two such criteria are needed to uniquely determine both coefficients~\cite{DunkelHaenggiWeber-2009-Phys.Rev.E79}, however, having set $\diffusion^2 / 2 = \diffusivity$ to a constant that can be numerically deduced by fitting the model to experimental data, one constraint will suffice in our case.
In an earlier version of the RDM, a linear approximation was used for the drift coefficient function that enabled an analytical solution of the Fokker--Planck equation \cite{Wolschin-1999-Eur.Phys.J.A5}.

\subsection{\label{sec:cgc}Expected stationary state}
The stochastic process defined in Eq.\,(\ref{eq:sde-rapidity}) would approach a stationary state if its time evolution continued past $t = \tf$.
We can estimate this state by assuming the formation of a color-glass condensate (CGC) \cite{GribovLevinRyskin-1983-Phys.Rep.100,MuellerQiu-1986-Nucl.Phys.B268,BlaizotMueller-1987-Nucl.Phys.B289,McLerranVenugopalan-1994-Phys.Rev.D49}, a coherent state based on the saturation of the gluon density below a characteristic momentum scale~$Q_{\mathrm{s}}$.
In the CGC framework, the post-collision distribution of the forward-moving participant protons is given by \cite{KharzeevKovchegovTuchin-2004-Phys.Lett.B599,BaierMehtarTaniSchiff-2006-Nucl.Phys.A764,DumitruHayashigakiJalilianMarian-2006-Nucl.Phys.A765}
\begin{equation}
	\pdfy\cgc(y) = \frac{C}{2\archimedesconstant} \int_0^1 \differential{x} \mathinner{q_v(x)} \mathinner{g(x^{2+\gssexponent} \eulersnumber^{\tau(y)})}
	\label{eq:pdf-cgc}%
	\mathinner{,}
\end{equation}
where $x$ is the longitudinal momentum fraction carried by the protons' valence quarks and $q_v$ denotes the valence-quark distribution function for which we use the NNLO results from \cite{MartinRobertsStirlingThorne-2002-Phys.Lett.B531}.
$C$ is a normalizing constant that sets the integral of $\pdfy\cgc$ to unity.
To determine the distribution function~$g$ of the soft gluons from the backward-moving fragment, we choose the Golec-Biernat--W{\"u}sthoff model \cite{GolecBiernatWuesthoff-1998-Phys.Rev.D59} in which $g$ reduces to a simple function of the scaling variable $\zeta = \bracks*{(p^1)^2 + (p^2)^2} / {Q_{\mathrm{s}}^2}$,
\begin{equation}
	g(\zeta) = 4\archimedesconstant \zeta \eulersnumber^{-\zeta}
	\mathinner{.}
\end{equation}

The gluon-saturation-scale exponent~$\gssexponent$ determines the $x$ dependence of $Q_{\mathrm{s}}$,
\begin{equation}
	Q_{\mathrm{s}}^2 = \gssprefactor A^{1/3} x^{-\gssexponent}
	\mathinner{,}
\end{equation}
while the constant $\gssprefactor$ sets its dimension and the mass number~$A$ its scaling with the nucleus's size.
Together with the center-of-mass energy per nucleon pair~$\sqrt{\sNN}$, the same three parameters determine the rapidity dependence of $\pdfy\cgc$ through the dimensionless function
\begin{equation}
	\tau(y) = \ln\parens*{\frac{\sNN}{\gssprefactor}} - \frac{1}{3} \ln(A) - 2 (1 + \gssexponent) y
	\mathinner{.}
\end{equation}

More details on the subject can be found in Refs.\,\cite{MehtarTaniWolschin-2009-Phys.Rev.Lett.102,MehtarTaniWolschin-2009-Phys.Rev.C80}, where similar distribution functions were fitted directly to proton-stopping data without considering a time evolution of the system.

The integral in Eq.\,(\ref{eq:pdf-cgc}) has no analytic solution, and hence, we solve it numerically with adaptive Gauss--Kronrod quadrature in all our computations.
For rapidities far from zero, however, analytical approximate solutions exist, which we will discuss briefly below.

For large positive rapidities, the argument of $g$ becomes very small such that we can approximate $g(\zeta) \approx 4\archimedesconstant \zeta$ and separate the $x$- and $y$-dependent terms,
\begin{equation}
	\pdfy\cgc(y) \underset{y \to +\infty}{\sim} 2 C \eulersnumber^{\tau(y)} \int_0^1 \differential{x} \mathinner{q_v(x)} x^{2+\gssexponent}
	\mathinner{.}
\end{equation}
With the integral yielding a constant numerical factor, the distribution function thus decays exponentially for large positive $y$, $\pdfy\cgc(y) = O\parens[\big]{\exp\parens*{\alpha_+ y}}$, with decay constant~$\alpha_+ = -2 (1 + \gssexponent)$.

For large negative rapidities, only small $x$ values contribute to the integral due to the exponential damping with $\tau(y)$.
If the low-$x$ behavior of the valence-quark distribution is given by $x \mathinner{q_v(x)} \sim a x^b$, Eq.\,(\ref{eq:pdf-cgc}) reduces to the definition of the gamma function times an exponential function of $\tau(y)$,
\begin{multline}
	\pdfy\cgc(y) \underset{y \to -\infty}{\sim} \frac{2 C a}{2 + \gssexponent} \mathinner{\Gamma\parens*{1 + \frac{b}{2 + \gssexponent}}}
	\\
	\times \mathinner{\exp\parens*{-\frac{b \mathinner{\tau(y)}}{2 + \gssexponent}}}
	\mathinner{.}
\end{multline}
Accordingly, the distribution function exhibits an exponential tail also for large negative values of $y$, where $\pdfy\cgc(y) = O\parens[\big]{\exp\parens*{\alpha_- y}}$ with $\alpha_- = 2 b (1 + \gssexponent) / (2 + \gssexponent)$.

\subsection{Fluctuation--dissipation relation}
A Fokker--Planck equation of the form Eq.\,(\ref{eq:fpe}) possesses a unique stationary solution~$\pdfy\stationary$.
It can be easily calculated by using the fact that its time derivative vanishes, $\partial_t \pdfy\stationary(y) = 0$, resulting in \cite{Ebeling-2000-Condens.MatterPhys.3,DunkelHilbertHaenggi-2006}
\begin{equation}
	\pdfy\stationary(y) \propto \exp\bracks*{\frac{1}{\diffusivity} \int_*^y \differential{y'} \drift(y')}
	\label{eq:pdf-stationary}%
	\mathinner{,}
\end{equation}
where the lower integration limit is chosen such that the integral exists.
All solutions of Eq.\,(\ref{eq:fpe}) would converge against this state for $t \to \infty$, $\lim_{t \to \infty} \pdfy(t;y) = \pdfy\stationary(y)$, if we continued their time evolution past $\tf$, which is, however, physically impossible since the fragments separate.
Hence, fixing the drift coefficient and diffusivity determines $\pdfy\stationary$ and \latin{vice versa}:
Inverting Eq.\,(\ref{eq:pdf-stationary}) yields the fluctuation--dissipation relation associated with a given stationary state~$\pdfy\stationary$ \cite{DebbaschMallickRivet-1997-J.Stat.Phys.88,DunkelHaenggiWeber-2009-Phys.Rev.E79,DunkelHaenggi-2009-Phys.Rep.471},
\begin{equation}
	\frac{\drift(y)}{\diffusivity} = \partial_y \ln\parens[\big]{\pdfy\stationary(y)}
	\label{eq:fdr}%
	\mathinner{.}
\end{equation}

\begin{figure}
	\includegraphics{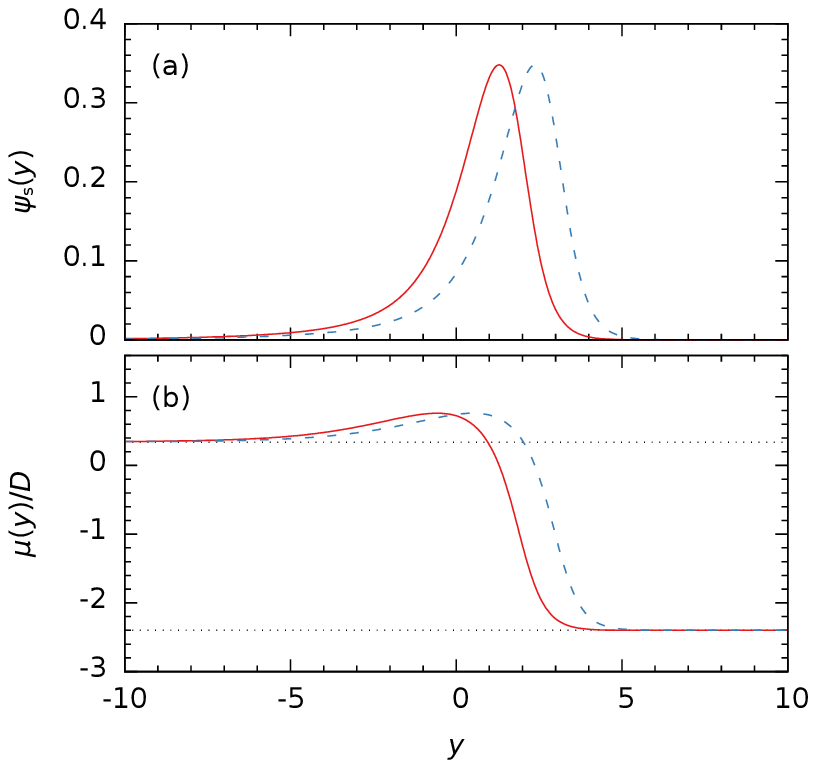}%
	\caption{\label{fig:pdf-stationary+fdr}%
		Stationary distribution function (a) and fluctuation--dissipation relation (b) for $\pdfy\stationary \equiv \pdfy\cgc$ of the forward-moving nucleus in a collision of \Pb~nuclei with center-of-mass energy $\sqrt{\sNN} = \SI{17.3}{\GeV}$ (solid, red) and \Au~nuclei with $\sqrt{\sNN} = \SI{62.4}{\GeV}$ (dashed, blue) with $\gssexponent = \num{0.2}$ and $\gssprefactor = \SI{0.09}{\GeV\squared}$.
		Dotted horizontal lines indicate the limiting values $\alpha_+ = \num{-2.4}$ and $\alpha_- \approx \num{+0,34}$.
		Decreasing $\gssexponent$ stretches the curves toward more positive $y$ and reduces their slope, while increasing $\gssprefactor/\sNN$ or $A$ shifts them to the left.
	}
\end{figure}

If we then identify $\pdfy\stationary \equiv \pdfy\cgc$ with the CGC distribution from Eq.\,(\ref{eq:pdf-cgc}), the drift coefficient~$\drift$ can thus be fixed as a function of $\diffusivity$ and $y$.
Like $\pdfy\cgc$, the resulting expression for $\drift$ is not analytic, but can be evaluated numerically as shown in Fig.\,\ref{fig:pdf-stationary+fdr}.
The graph is roughly \textsf{S}-shaped and converges toward constant values for $y \to \pm\infty$ due to the exponential tails of $\pdfy\cgc$,
\begin{subequations}
\begin{align}
	\lim_{y \to +\infty} \frac{\drift(y)}{\diffusivity} &= \alpha_+ = -2 (1 + \gssexponent)
	\mathinner{,}
	\\
	\lim_{y \to -\infty} \frac{\drift(y)}{\diffusivity} &= \alpha_- = +2 b \mathinner{\frac{1 + \gssexponent}{2 + \gssexponent}}
	\mathinner{.}
\end{align}
\end{subequations}
Its zero crossing marks the peak position of $\pdfy\cgc$; the maximum close to $y \approx 0$ indicates an inflection point of the logarithm of $\pdfy\cgc$.

\section{\label{sec:results}Results}
We obtain a dimensionless form of the Fokker--Planck equation~(\ref{eq:fpe}) by substituting the time~$t$ with the evolution parameter $s(t) = (t - \ti) / (\tf - \ti)$ and reordering some terms.
The transformed equation reads
\begin{equation}
	\partial_s \pdfy\parens[\big]{t(s);y} = \evolutionscale \bracks*{-\partial_y \frac{\drift(y)}{\diffusivity} + \partial_y^2} \pdfy\parens[\big]{t(s);y}
	\label{eq:fpe-dimensionless}%
\end{equation}
with $\timespan = \tf - \ti$.
While $s$, $y$, and $\pdfy$ are dimensionless by definition, we have arranged the remaining quantities such that they form the composite dimensionless factors $\evolutionscale$ and $\drift(y) / \diffusivity$.
The latter is given by the fluctuation--dissipation relation defined in Eq.\,(\ref{eq:fdr}), while $\evolutionscale$ is treated as a free parameter of the model.

As the strength of the stochastic processes scale with the diffusivity~$\diffusivity$, while $\timespan$ is defined as the time span during which the system is subject to the associated forces, the compound variable $\evolutionscale$ can be interpreted as the net impact of the partonic interactions between the nuclei.
Appearing only on the right-hand side of Eq.\,(\ref{eq:fpe-dimensionless}), it can be completely absorbed into the evolution parameter by rescaling $\tilde{s} = \evolutionscale \times s$, which then runs from $\tilde{s}(\ti) = 0$ to $\tilde{s}(\tf) = \evolutionscale$.
Small values of $\evolutionscale$ hence indicate that the system remains close to its initial state, while larger values drive it closer toward the stationary state imposed by the fluctuation--dissipation relation.

The transformed Fokker--Planck equation~(\ref{eq:fpe-dimensionless}) is solved numerically for $0 < s \leq 1$ by discretizing the rapidity derivative operators and solving the resulting system of ordinary differential equations with an additive Runge--Kutta method~\cite{KennedyCarpenter-2003-Appl.Numer.Math.44}.

\begin{table*}
	\caption{\label{tab:parameters}%
		Parameters used in the model as determined through a fit of the final net-proton distribution functions to experimental data~\cite{Appelshaeuser-1999-Phys.Rev.Lett.82,Bearden-2004-Phys.Rev.Lett.93,Debbe-2008-J.Phys.GNucl.Part.Phys.35,Arsene-2009-Phys.Lett.B677}.
		For $\sqrt{\sNN} = \SI{200}{\GeV}$, final and stationary state were so close to each other that no meaningful fit result could be determined for $\evolutionscale$ (see text); accordingly, no numerical value is given at this energy.
		$\gssexponent$ and $\gssprefactor$ are shared parameters and hence take the same numerical values for all collisions.
		Reduced sums of squared residuals~$\chi^2 / \ndf$ (excluding shared parameters) are given for each setting as measures for the individual goodness-of-fit.
	}
	\begin{ruledtabular}
		\begin{tabular}{r D{.}{.}{3,1} D{.}{.}{1,2} D{-}{\text{--}}{1,2} D{.}{.}{1,1} D{.}{.}{1,2} D{.}{.}{3,0} D{.}{.}{1,1} D{.}{.}{1,2}}
			Nuclei & \multicolumn{1}{c}{$\sqrt{\sNN}$ (\si{\GeV})} & \multicolumn{1}{c}{$y\beam$} & \multicolumn{1}{c}{Centrality (\si{\percent})} & \multicolumn{1}{c}{$\gssexponent$} & \multicolumn{1}{c}{$\gssprefactor$ (\si{\GeV\squared})} & \multicolumn{1}{c}{$\N$} & \multicolumn{1}{c}{$\evolutionscale$} & \multicolumn{1}{c}{$\chi^2 / \ndf$} \\
			\noalign{\vskip+0.5ex}\colrule\noalign{\vskip+1ex}
			\Pb &  17.3 & 2.909 & 0-5  & 0.2 & 0.09 & 150 & 3.1 & 0.56 \\
			\Au &  62.4 & 4.196 & 0-10 & 0.2 & 0.09 & 140 & 3.8 & 1.4  \\
			\Au & 200   & 5.361 & 0-5  & 0.2 & 0.09 & 150 & \multicolumn{1}{c}{---} & 0.36 \\
			\Au & 200   & 5.361 & 0-10 & 0.2 & 0.09 & 120 & \multicolumn{1}{c}{---} & 1.4  \\
		\end{tabular}
	\end{ruledtabular}
\end{table*}

\begin{figure}
	\includegraphics{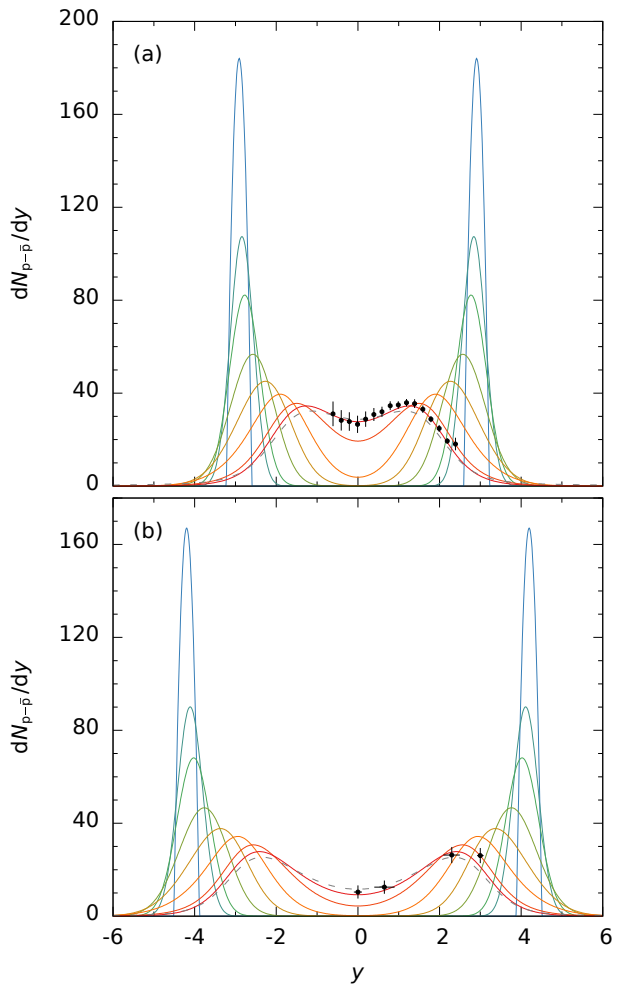}%
	\caption{\label{fig:results-low-cme}%
		Time evolution of the net-proton rapidity distribution function for central collisions of \Pb~nuclei with center-of-mass energy $\sqrt{\sNN} = \SI{17.3}{\GeV}$ at \SIrange{0}{5}{\percent} centrality (a) and \Au~nuclei with $\sqrt{\sNN} = \SI{62.4}{\GeV}$ at \SIrange{0}{10}{\percent} centrality (b).
		Solid lines mark the time steps $s = \numlist{0;0.01;0.02;0.05;0.1;0.2;0.5;1}$, where $s = 0$ corresponds to the initial state (peaked, blue) and $s = 1$ to the final state (broad, red).
		The latter is compared with experimental data (black circles) recorded at SPS by the NA49 Collaboration~\cite{Appelshaeuser-1999-Phys.Rev.Lett.82} (a) and RHIC by the BRAHMS Collaboration~\cite{Arsene-2009-Phys.Lett.B677} (b); associated uncertainties are depicted as bars.
		Dashed lines indicate the stationary distribution functions ($s \to \infty$).
	}
\end{figure}

\begin{figure}
	\includegraphics{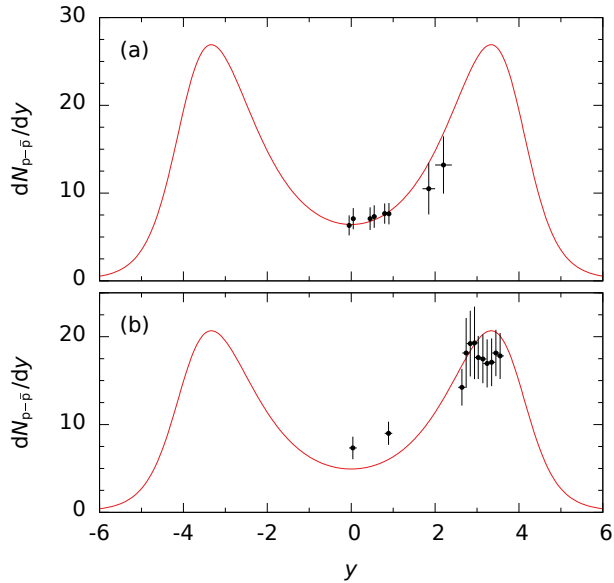}%
	\caption{\label{fig:results-high-cme}%
		Calculated stationary net-proton rapidity distribution functions for collisions of \Au~nuclei with center-of-mass energy $\sqrt{\sNN} = \SI{200}{\GeV}$ at \SIrange{0}{5}{\percent} centrality (a) and \SIrange{0}{10}{\percent} centrality (b).
		Black circles show experimental data from RHIC recorded by the BRAHMS Collaboration in 2004~\cite{Bearden-2004-Phys.Rev.Lett.93} and 2008~\cite{Debbe-2008-J.Phys.GNucl.Part.Phys.35}, respectively; uncertainties are depicted as bars.
	}
\end{figure}

We compare the results of our calculations to SPS and RHIC data from the NA49 and BRAHMS Collaboration, respectively \cite{Appelshaeuser-1999-Phys.Rev.Lett.82,Bearden-2004-Phys.Rev.Lett.93,Debbe-2008-J.Phys.GNucl.Part.Phys.35,Arsene-2009-Phys.Lett.B677}.
The gluon-saturation-scale exponent~$\gssexponent$ and prefactor~$\gssprefactor$, the net-proton number $\N$, and the diffusivity times elapsed time~$\evolutionscale$ are free parameters of the model.
They are determined through a simultaneous weighted least-squares fit of the final net-proton distribution functions to the experimental data, where minimization of the fit objective is done numerically with a
quasi-Newton
method~\cite{Nocedal-1980-Math.Comput.35,LiuNocedal-1989-Math.Program.45}.
We restrict $\N$ to deviate not more than \SI{10}{\percent} from the respective Glauber result, while $\gssexponent$ and $\gssprefactor$ are treated as common parameters that take the same numerical values for all collisions in the SPS to RHIC energy region.
Our results are given in Tab.\,\ref{tab:parameters}; the combined sum of squared residuals divided by the total number of degrees of freedom is $\chi^2 / \ndf \approx \num{0.89}$.
The estimates for $\gssexponent$ and $\gssprefactor$ compare well to literature results, where $\gssexponent \approx \num{0.288}$ and $\gssprefactor \approx \SI{0.097}{\GeV\squared}$ were obtained in a fit to deep-inelastic-scattering data from the DESY Hadron--Electron Ring Accelerator (HERA) \cite{GolecBiernatWuesthoff-1998-Phys.Rev.D59}.

The time evolution of the net-proton distribution functions for the two collisions with lower energy is shown in Fig.\,\ref{fig:results-low-cme}.
As expected \cite{JiShenYi-2018-J.Dyn.Diff.Equat.31}, the distribution functions converge exponentially in time toward the stationary state, which is indicated in the plot by a logarithmic spacing of the intermediate time steps.
While the final distribution functions appear to differ only slightly from the stationary ones, the systems are still far from their stationary states in a temporal sense, as further convergence slows down exponentially.

At the lower center-of-mass energies $\sqrt{\sNN} = \SIlist{17.3;62.4}{\GeV}$, the final and stationary state differ enough for a reasonable estimate of $\evolutionscale$, which takes values between \num{3} and \num{4}.
At the higher energy $\sqrt{\sNN} = \SI{200}{\GeV}$, however, final and stationary state are too close compared to the experimental errors.
As a consequence of the exponential convergence in time, the uncertainty in the determination of $\evolutionscale$ becomes orders of magnitude larger than the actual value.
Therefore, a meaningful estimate of $\evolutionscale$ is not possible and no values are given in Tab.\,\ref{tab:parameters} at this center-of-mass energy.

Fig.\,\ref{fig:results-high-cme} therefore shows only the stationary net-proton distribution functions for the two collisions at $\sqrt{\sNN} = \SI{200}{\GeV}$, which are nearly indistinguishable from the final distribution functions.
A time evolution from the initial to the final state cannot be given due to the indeterminate value of $\evolutionscale$.
The net-proton numbers differ for the two centralities, being higher for \SIrange{0}{5}{\percent} and lower for \SIrange{0}{10}{\percent}, which is consistent with the latter data containing additional events with fewer participants.



All required numerical routines were implemented with the \computerprogram{Julia} programming language~\cite{BezansonEdelmanKarpinskiShah-2017-SIAMRev.59}; functionalities for the solution of differential equations and parameter optimization were provided by the packages \computerprogram{DifferentialEquations.jl}~\cite{RackauckasNie-2017-J.OpenRes.Softw.5} and \computerprogram{Optim.jl}~\cite{MogensenRiseth-2018-J.OpenSourceSoftw.3}, respectively.

\section{\label{sec:conclusions}Conclusions}
A relativistic phase-space diffusion model for the time evolution of net-proton distribution functions in rapidity space was presented to account for the transition process from the initial to the final state in baryon stopping.
Inspired by the phenomenological RDM, the model uses similar key assumptions, but is based on stochastic particle trajectories constructed from relativistic Markov processes in phase space that are equivalent to non-Markovian spacetime processes.
The drift and diffusion coefficient, which carry over from the Langevin to the Fokker--Planck formulation of the system's time evolution, were determined by assuming a constant diffusivity in rapidity space and setting the stationary solution of the Fokker--Planck equation to a QCD-inspired distribution function.
Due to the latter's exponential tails, the associated fluctuation--dissipation relation was found to be virtually constant for large absolute rapidities.
Analytic expressions for the limiting values were derived.

A simultaneous least-squares fit was used to determine the free model parameters
for four data sets recorded at SPS and RHIC by the NA49 and BRAHMS Collaboration, respectively.
In the fit, the net-proton number~$\N$ was restricted to a neighborhood of the corresponding Glauber result for each collision.
The gluon-saturation-scale exponent~$\gssexponent$ and prefactor~$\gssprefactor$ were treated as common parameters taking the same value in all comparisons with experiment.
No constraints, apart from positivity, were placed on the dimensionless factor~$\evolutionscale$ composed of the diffusivity and the elapsed time between initial and final state.
For {\Pb} and {\Au} collisions at $\sqrt{\sNN} = \SIlist{17.3;62.4}{\GeV}$, respectively, agreement with the data could be reached and an estimate of the time evolution from the initial to the final state was given.
At \SI{200}{\GeV}, the latter was not possible, since the final and stationary distribution functions were found to be too close compared to the experimental uncertainties.


The phase-space diffusion framework adopted in this article is easily adaptable to different physical systems and allows to construct the drift and diffusion coefficient functions, which can be difficult to access theoretically, from mesoscopic considerations.
In a forthcoming work, we will examine a possible application to charged-particle production in relativistic heavy-ion collisions.

\begin{acknowledgments}
	J.\,H.~acknowledges support from a research scholarship of the Heidelberg Graduate School for Physics and from a graduate scholarship of the Villigst Scholarship Foundation.
\end{acknowledgments}

\bibliography{references}

\end{document}